\documentclass[12pt]{article}
\usepackage{epsfig}
\usepackage{feynmf}
\usepackage{a4p}
\usepackage{cite}

\newcommand{\sqee}{\sqrt{s_{\rm ee}}}

\def\gg{\gamma\gamma}
\def\ee{\mbox{e}^+\mbox{e}^-}

\def\sqee{\sqrt{s}_{\rm ee}}
\def\see{s_{\rm ee}}

\def\pt{p_{\rm T}}
\def\ptj{p_{\rm T}^{{\rm J}/\psi}}

\def\ppbar{\mbox{p}\overline{\mbox{p}}}

\def\pz{\phantom{0}}
\def\pzz{\phantom{00}}
\def\Chi{\chi_{\rm c2}}
\def\Gam{\Gamma(\chi_{\rm c2} \to \gg)}
\def\DM{M_{\ell\ell\gamma}-M_{\ell\ell}}
\def\BR{\mbox{BR}}
\def\result{1.76 \pm 0.47 \pm 0.37 \pm 0.15}
\begin{document}
\renewcommand{\thefootnote}{\arabic{footnote}}

\begin{titlepage}
\begin{center}{\large   EUROPEAN LABORATORY FOR PARTICLE PHYSICS
}\end{center}\bigskip
\begin{flushright}
       CERN-EP/98-106   \\ 3 July 1998
\end{flushright}
\bigskip\bigskip\bigskip\bigskip\bigskip
\begin{center}{\LARGE\bf
\boldmath Production of $\Chi$ Mesons

in Photon-Photon Collisions at LEP\unboldmath
}\end{center}\bigskip\bigskip
\begin{center}{\LARGE The OPAL Collaboration
}\end{center}\bigskip\bigskip
\begin{center}
\end{center}
\vspace{0.5cm}
\begin{center}
\end{center}
\bigskip\begin{center}{\large  Abstract}\end{center}
We present an observation at LEP of the production
of $\chi_{\rm c2}$ mesons in the collisions
of two quasi-real photons using the OPAL detector.
The $\chi_{\rm c2}$ mesons are reconstructed in the decay channel
$\Chi \to  J/\psi ~\gamma \to  \ell^+~ \ell^- ~\gamma $ (with $\ell$ = $e, \mu$)
using all data taken at $\ee$ centre-of-mass energies
of $91$ and $183$~GeV, corresponding to integrated luminosities of
167 and 55 pb$^{-1}$ respectively.
The two-photon width of the $\Chi$ is determined to be
$$\Gam=\result\mbox{~keV} ,$$
where the first error is statistical, the second is systematic and
the third comes from branching ratio uncertainties.
\bigskip\bigskip\bigskip\bigskip
\bigskip\bigskip
\begin{center}{\large
(To be submitted to Physics Letters)}
\end{center}
\end{titlepage}
\begin{center}{\Large        The OPAL Collaboration
}\end{center}\bigskip
\begin{center}{
K.\thinspace Ackerstaff$^{  8}$,
G.\thinspace Alexander$^{ 23}$,
J.\thinspace Allison$^{ 16}$,
N.\thinspace Altekamp$^{  5}$,
K.J.\thinspace Anderson$^{  9}$,
S.\thinspace Anderson$^{ 12}$,
S.\thinspace Arcelli$^{  2}$,
S.\thinspace Asai$^{ 24}$,
S.F.\thinspace Ashby$^{  1}$,
D.\thinspace Axen$^{ 29}$
% ...
}\end{center}\bigskip
\bigskip
$^{  1}$School of Physics and Astronomy, University of Birmingham,
Birmingham B15 2TT, UK
\newline
$^{  2}$Dipartimento di Fisica dell' Universit\`a di Bologna and INFN,
I-40126 Bologna, Italy
\newline
$^{  3}$Physikalisches Institut, Universit\"at Bonn,
D-53115 Bonn, Germany
\newline
$^{  4}$Department of Physics, University of California,
Riverside CA 92521, USA
\newline
$^{  5}$Cavendish Laboratory, Cambridge CB3 0HE, UK
\newline
$^{  6}$Ottawa-Carleton Institute for Physics,
Department of Physics, Carleton University,
Ottawa, Ontario K1S 5B6, Canada
\newline
$^{  7}$Centre for Research in Particle Physics,
Carleton University, Ottawa, Ontario K1S 5B6, Canada
\newline
$^{  8}$CERN, European Organisation for Particle Physics,
CH-1211 Geneva 23, Switzerland
\newline
$^{  9}$Enrico Fermi Institute and Department of Physics,
University of Chicago, Chicago IL 60637, USA
\newline
$^{ 10}$Fakult\"at f\"ur Physik, Albert Ludwigs Universit\"at,
D-79104 Freiburg, Germany
\newline
$^{ 11}$Physikalisches Institut, Universit\"at
Heidelberg, D-69120 Heidelberg, Germany
\newline
$^{ 12}$Indiana University, Department of Physics,
Swain Hall West 117, Bloomington IN 47405, USA
\newline
$^{ 13}$Queen Mary and Westfield College, University of London,
London E1 4NS, UK
\newline
$^{ 14}$Technische Hochschule Aachen, III Physikalisches Institut,
Sommerfeldstrasse 26-28, D-52056 Aachen, Germany
\newline
$^{ 15}$University College London, London WC1E 6BT, UK
\newline
$^{ 16}$Department of Physics, Schuster Laboratory, The University,
Manchester M13 9PL, UK
\newline
$^{ 17}$Department of Physics, University of Maryland,
College Park, MD 20742, USA
\newline
$^{ 18}$Laboratoire de Physique Nucl\'eaire, Universit\'e de Montr\'eal,
Montr\'eal, Quebec H3C 3J7, Canada
\newline
$^{ 19}$University of Oregon, Department of Physics, Eugene
OR 97403, USA
\newline
$^{ 20}$CLRC Rutherford Appleton Laboratory, Chilton,
Didcot, Oxfordshire OX11 0QX, UK
\newline
$^{ 22}$Department of Physics, Technion-Israel Institute of
Technology, Haifa 32000, Israel
\newline
$^{ 23}$Department of Physics and Astronomy, Tel Aviv University,
Tel Aviv 69978, Israel
\newline
$^{ 24}$International Centre for Elementary Particle Physics and
Department of Physics, University of Tokyo, Tokyo 113, and
Kobe University, Kobe 657, Japan
\newline
$^{ 25}$Institute of Physical and Environmental Sciences,
Brunel University, Uxbridge, Middlesex UB8 3PH, UK
\newline
$^{ 26}$Particle Physics Department, Weizmann Institute of Science,
Rehovot 76100, Israel
\newline
$^{ 27}$Universit\"at Hamburg/DESY, II Institut f\"ur Experimental
Physik, Notkestrasse 85, D-22607 Hamburg, Germany
\newline
$^{ 28}$University of Victoria, Department of Physics, P O Box 3055,
Victoria BC V8W 3P6, Canada
\newline
$^{ 29}$University of British Columbia, Department of Physics,
Vancouver BC V6T 1Z1, Canada
\newline
$^{ 30}$University of Alberta,  Department of Physics,
Edmonton AB T6G 2J1, Canada
\newline
$^{ 31}$Duke University, Dept of Physics,
Durham, NC 27708-0305, USA
\newline
$^{ 32}$Research Institute for Particle and Nuclear Physics,
H-1525 Budapest, P O  Box 49, Hungary
\newline
$^{ 33}$Institute of Nuclear Research,
H-4001 Debrecen, P O  Box 51, Hungary
\newline
$^{ 34}$Ludwigs-Maximilians-Universit\"at M\"unchen,
Sektion Physik, Am Coulombwall 1, D-85748 Garching, Germany
\newline
\bigskip\newline
$^{  a}$ and at TRIUMF, Vancouver, Canada V6T 2A3
\newline
$^{  b}$ and Royal Society University Research Fellow
\newline
$^{  c}$ and Institute of Nuclear Research, Debrecen, Hungary
\newline
$^{  d}$ and Department of Experimental Physics, Lajos Kossuth
University, Debrecen, Hungary
\newline
$^{  e}$ on leave of absence from the University of Freiburg
\newline

\section{Introduction}
We present a LEP measurement of the two-photon width $\Gam$
for the production of $\Chi$ mesons in photon-photon interactions.
The $\Chi$ is observed through its decay to a $J/\psi$ and a photon, 
with the $J/\psi$ identified through its decay to
$\ee$ or $\mu^+ \mu^-$. 
All LEP1 and LEP2 data samples taken
at centre-of-mass energies $\sqee$ of $91$ and $183$~GeV are used,
corresponding to integrated luminosities $L=167$ and $55$~pb$^{-1}$,
respectively.

The $\Chi$ meson is a $\rm c\bar{c}$ bound state with 
spin-parity $J^{\rm PC} = 2^{++}$ which can therefore be produced in the 
collision of two photons. At LEP,
virtual photons (denoted by $\gamma^{\ast}$)
are emitted by the beam electrons\footnote{Positrons
are also referred to as electrons}.
The $\Chi$ mesons can therefore be produced in the process 
$\ee \to \ee \gamma^{\ast} \gamma^{\ast} \to \ee \Chi $.
Since the photons carry only
small negative squared four-momenta $Q^2$, they are assumed to 
be quasi-real ($Q^2 \approx 0$). 

The two-photon width $\Gamma_{\gamma\gamma}$ measures
the coupling of two real photons to a resonance and is directly 
proportional to the production cross-section in $\gg$ collisions.
Due to the relatively large mass $m_{\rm c}$ of the c quarks 
the $\Chi$ meson can be treated non-relativistically and perturbatively. 
The ratio of the widths of the $\Chi$ decaying to two photons and to
two gluons can be directly related to the ratio of the
electromagnetic and strong coupling constants:
\begin{equation}
\frac{\Gam}{\Gamma(\Chi \to {\rm gg})}
= \frac{8}{9} \left(\frac{\alpha_{\rm em}}{\alpha_{\rm s}(m_{\rm c})}
   \right)^2.
\end{equation}
In simple models 
$\Gamma(\Chi \to {\rm gg})\approx\Gamma(\Chi \to \rm hadrons)$.
Using $\Gamma(\Chi \to \rm hadrons)$ measured in $\ppbar$ 
scattering~\cite{bib-ppbar} and  
$\alpha_{\rm s}(m_{\rm c})= 0.3$,
we expect $\Gam$ to be of the order of 1 keV. 
Based on a QCD model of Bodwin et~al.~\cite{bodwin}
a value of $\Gam = 0.82\pm0.30$~keV is obtained. 
A recent calculation by Schuler~\cite{schuler}
predicts a smaller
value for $\Gam$ of 0.28 keV.
The experimental results on the two-photon width $\Gam$ measured
in $\ee$~\cite{bib-cleo,bib-tpc} and $\ppbar$ 
collisions~\cite{bib-E760b,bib-r704}
are summarised in Table~\ref{tab-results}.
The measurements vary over an order of magnitude.
The result $0.321\pm0.095$~keV, obtained by E760~\cite{bib-E760b} from
the process $\ppbar\to\Chi\to\gamma\gamma$, dominates the current 
world average of $0.37\pm0.17$~keV~\cite{bib-pdg}.

\section{The OPAL detector}
A detailed description of the OPAL detector
can be found in Ref.~\cite{opaltechnicalpaper}, and
therefore only a brief account of the main features relevant
to the present analysis will be given here.
 
The central tracking system is located inside 
a solenoidal magnet which
provides a uniform axial magnetic field of 0.435~T along the beam
axis\footnote{In the OPAL coordinate system 
the $z$ axis points in the direction of the e$^-$ beam. The
polar angle $\theta$, the azimuthal angle $\phi$
and the radius $r$ denote the usual spherical coordinates.}.
The detection efficiency for charged particles is close to 100~$\%$
within the polar angle range $|\cos\theta|<0.92$.
The magnet is surrounded in the barrel region ($|\cos\theta|<0.82$)
by a lead glass electromagnetic
calorimeter (ECAL) and a hadronic sampling calorimeter (HCAL).  
Outside the HCAL, the detector is surrounded by muon
chambers. There are similar layers of detectors in the 
endcaps ($0.81<|\cos\theta|<0.98$). 
The small angle region from 47 to 140 mrad
around the beam pipe on both sides
of the interaction point is covered by the forward calorimeters (FD)
and the region from 25 to 59 mrad by the silicon tungsten luminometers (SW).
From 1996 onwards, which includes the data-taking period at $\sqee=183$~GeV
presented in this paper,
the lower boundary of the acceptance has been  increased to 33 mrad
following the installation of a low angle shield to protect the
central detector against possible synchrotron radiation.
 
Starting with the innermost components, the
tracking system consists of a high precision silicon
microvertex detector, a vertex
drift chamber, a large-volume jet chamber with 159 layers of axial
anode wires and a set of $z$ chambers measuring the track coordinates
along the beam direction. 
The transverse momenta $\pt$ of tracks with respect to the $z$ axis
are measured with a precision 
parametrised by
$\sigma_{\pt}/\pt=\sqrt{0.02^2+(0.0015\cdot \pt)^2}$ ($\pt$ in GeV/$c$)
in the central region. 
The jet chamber also provides 
measurements of the energy loss, ${ \rm d} E/ {\rm d}x$, 
which are used for particle identification~\cite{opaltechnicalpaper}.

The barrel and endcap sections of the ECAL  are
both constructed from lead glass blocks with a depth of
$24.6$ radiation lengths in the barrel region and more than 
$22$ radiation lengths in the endcaps. 
The FD consist of cylindrical lead-scintillator calorimeters with a depth of   
24 radiation lengths divided azimuthally into 16 segments.  
The electromagnetic energy resolution is about
$18\%/\sqrt{E}$, where $E$ is in GeV.                                  
The SW detectors~\cite{bib-siw} consist
of 19 layers of silicon and 18
layers of tungsten, corresponding to a total of 22 radiation
lengths. Each silicon layer consists of 16 wedge
shaped silicon detectors. The electromagnetic energy resolution is about
$25\%/\sqrt{E}$ ($E$ in GeV).

\section{Process kinematics and Monte Carlo simulation}
\label{sec-MC}
The $\ee$ cross-section 
for the production of $\Chi$ mesons in two-photon events
can  be described by the product of the $\gg$ luminosity function 
$\cal{L}_{\gg}$ and the
cross-section $\sigma$ of the process $\gamma^{\ast} \gamma^{\ast} \to \Chi$:
\begin{equation}
 \sigma(\ee \to \ee \Chi) = \cal{L}_{\gg}(\ee \to \ee \gamma^{\ast} 
\gamma^{\ast})
   \sigma(\gamma^{\ast} \gamma^{\ast} \to \Chi).
\end{equation}
Since most photons have a small negative squared four-momentum-transfer 
$Q^2$, the 
photons are considered to be quasi-real with helicity $\pm 1$,
i.e. they are transverse.
Thus only the luminosity function describing two transverse 
photons contributes. 
The two-photon formation cross-section for the $\Chi$ resonance  
with this assumption~\cite{bib-budnev} is
\begin{equation}
\sigma(\gamma^{\ast} \gamma^{\ast} \to \Chi) = 8 \pi
       (2J_{\Chi}+1) \Gam 
       \frac{\Gamma_{\rm tot}}
       {(W^2-M_{\Chi}^2)^2 + M_{\Chi}^2 \Gamma^2_{\rm tot}},
\label{eq-sgg}
\end{equation}
where $J_{\Chi}$ denotes the spin of the $\Chi$, $M_{\Chi}$ its mass, 
$\Gamma_{\rm tot}$ its total width and $W$ the invariant mass of the 
two-photon system.
The total cross-section for
the process $ \ee \to \ee \Chi$ can be written as~\cite{bib-budnev}:
\begin{equation}
\sigma(\ee \to \ee \Chi) = (2J_{\Chi}+1)\frac{8\alpha^2_{\rm em} 
\Gam}{M_{\Chi}^3} g(M^2_{\Chi}/\see ),
\label{eq-ggres}
\end{equation}
where the dependence on the squared $\ee$ centre-of-mass energy $\see$
is contained in the function
\begin{equation}
  g(z)=
\left(\left(1+\frac{1}{2}z\right)^2 \ln\frac{1}{z} - \frac{1}{2}
         (1-z)(3+z)\right)
\left(\ln \frac{M_{\rm J/\psi}^2}
      {z M_{\rm e}^2}-1\right)^2 -
      \frac{1}{3} \left(\ln\frac{1}{z}\right)^3.
\label{eq-fz}
\end{equation}
This expression is obtained by integrating
the luminosity function based on the equivalent photon approximation (EPA). 
Further parameters are the electron mass $M_{\rm e}$ and 
the mass $M_{\rm J/\psi}$ of the $J/\psi$ meson 
that enters into the form factor.
By measuring $\sigma(\ee\to \ee \Chi)$ the two-photon width $\Gam$
can be calculated.

The Monte Carlo Generator TWOGEN~\cite{bib-twogen} is used to generate 
the process $\gamma^{\ast}\gamma^{\ast}\to\Chi$ according to Eq.~\ref{eq-sgg}.
The decay of the resonance is handled by JETSET~7.408~\cite{bib-jetset} 
taking into account final state radiation of the two decay leptons.
The polar angles of the scattered electrons in
the process $\ee \to \ee \gamma^{\ast} \gamma^{\ast}$ are allowed to vary
from 0 to $\pi$. The $Q^2$ dependence of the $\Chi$ production cross-section
is modelled with a $J/\psi$ form factor.
All Monte Carlo events were generated
with a full simulation of the OPAL detector~\cite{bib-gopal}.
 
Only helicity $\pm2$ is relevant for the $\Chi$ production
amplitude~\cite{bib-close} where the helicity is defined with respect to
the $\Chi$ direction of motion in the laboratory system which
is collinear with the axis of the incoming photons.
In the decay of the $\Chi$ the photon carries an angular momentum of 
1, 2 or 3 (in units of $\hbar$). Since the $\Chi$ and the $J/\psi$ have 
opposite parity,
electric dipole (E1), magnetic quadrupole (M2) and electric octupole (E3) 
transitions
are possible. In terms of helicities, decay amplitudes for
helicity 0, $\pm1$ and $\pm2$
parametrise the dynamics of the $\Chi$ decay process. These decay helicity 
amplitudes
are defined with respect to the $J/\psi~\gamma$ axis and
are related to the angular momentum carried by the photon.

Three angles are necessary to describe the
$\Chi$ decay to a $J/\psi$ and a $\gamma$ with the
subsequent decay of the $J/\psi$ to a lepton pair.  
These angles are illustrated in Fig.~\ref{fig-angles}.
The polar angle $\theta^{\ast}$ is measured in the $\Chi$ rest frame 
and is the angle between the
$\Chi$ direction of motion in the laboratory system
and the outgoing decay photon. 
The polar and azimuthal angles of the positive decay lepton, $\theta^{'}$
and $\phi^{'}$, 
are measured in the $J/\psi$ rest frame with the $z'$ axis 
given by the direction opposite to
the photon's direction from the $\Chi$ decay,
and the $x'$ axis in the plane 
containing the $\Chi$ direction and the incident photons.

In TWOGEN the events are generated flat in $\cos\theta^{\ast}$, 
$\cos\theta^{'}$ and $\phi^{'}$. 
The Monte Carlo events are re-weighted according to the appropriate angular 
distribution of the $\Chi\to J/\psi~\gamma$ decay.
The full expression for the angular distribution is given 
in the Appendix.  
All decay amplitudes for helicities 0, $\pm1$ and $\pm2$ are taken into account.

We also used
the Monte Carlo generator GALUGA~\cite{bib-galuga} to  
generate $\Chi$ events. The selection efficiency for $\Chi$ events is
found to be consistent  with the selection efficiency
determined with TWOGEN. In addition, GALUGA is used to
generate the possible resonant background process 
$\ee\to\ee\chi_{\rm c1}$.

\section{Event selection}
\label{sec-evsel}
The production of $\Chi$ mesons in two-photon events is studied using
the data taken at centre-of-mass energies $\sqee$
of $91$ and $183$~GeV. 
The same set of cuts for both centre-of-mass energies is applied to the data 
to select $\Chi$ candidate events: 

\begin{itemize}
\item The sum of all energy deposits has to be less than 30 GeV 
      in the ECAL and
      less than 20 GeV in the HCAL. These cuts mainly reject
      two-fermion events.
\item Tracks are required to have at least 20 hits in the central jet chamber 
      used for the determination of the specific energy 
      loss ${\rm d}E/{\rm d}x$.
      Events are required to contain exactly two oppositely charged tracks,
      where the two tracks must have a minimum transverse momentum $\pt$
      with respect to the $z$ axis of the detector
      of $800$~MeV/$c$. This reduces the background
      from leptonic two-photon events with a small invariant mass $W$.
      The distance of the point of closest 
      approach to the origin of the tracks 
      in the $r\phi$ plane must be less than 
      5~cm in the $z$ direction 
      and less than 1 cm in the $r\phi$ plane. 
      The invariant mass of the two tracks must lie in the range 
      between 2.7 and  3.5 GeV$/c^2$, close to the mass of the $J/\psi$.
      In order to ensure that
      the final state consists of only two tracks, 
      no additional track with $\pt>120$~MeV/$c$ 
      and more than 20 ${\rm d}E/{\rm d}x$ hits is allowed.
\item After these cuts the main background contribution is expected
      to be leptonic two-photon events, $\ee\to\ee\ee$ and
      $\ee\to\ee\mu^+\mu^-$.
      To reduce this background,
      the transverse momentum with respect to 
      the $z$ axis, $\ptj$, of the $J/\psi$ candidate formed
      by the two tracks is required to be larger than 100 MeV/$c$.
\item For the case where the $J/\psi$ decays to a muon (electron) pair, 
      the two tracks are identified as muons (electrons) 
      if the ${\mathrm d}E/{\mathrm d}x$ probability for the muon
      (electron) hypothesis  exceeds 1\%.
\item The event is rejected
      if a cluster with an energy deposit larger than 1~GeV
      is observed in the FD or SW. 
\item There must be exactly one photon candidate in an event.
      A cluster in the ECAL is treated as a photon candidate if it fulfils the
      following requirements: 
\begin{itemize}
\item 
      The measured cluster energy, $E_{\gamma}^{\rm meas}$,
      should be at least 300 MeV. 
      In order to reduce background from ECAL clusters 
      in the regions close to the beam pipe and
      in the overlap regions between the endcaps and the barrel,
      clusters with $|\cos\theta_{\gamma}| > 0.96$ and clusters with 
      \mbox{$0.84  < |\cos\theta_{\gamma}| <  0.85$} are not considered where 
      $\theta_{\gamma}$ is the polar angle of the cluster with 
      respect to the $z$ axis. 
\item
      The angles between the cluster and each track momentum 
      should exceed 300 mrad to reject events with final-state radiation in
      leptonic two-photon events.  
\end{itemize}      
\item The angle in the $r\phi$ plane between the photon and 
the $J/\psi$ candidate formed by the two tracks must be greater than
$150^{\circ}$. 
\end{itemize}
After these preselection cuts, 203 events remain at $\sqee=91$~GeV
and 53 events at $\sqee=183$~GeV.

\section{\boldmath $\Chi$ \unboldmath reconstruction and results}
\begin{figure}[htbp]
   \begin{center}
      \mbox{
          \epsfxsize=17.cm
          \epsffile{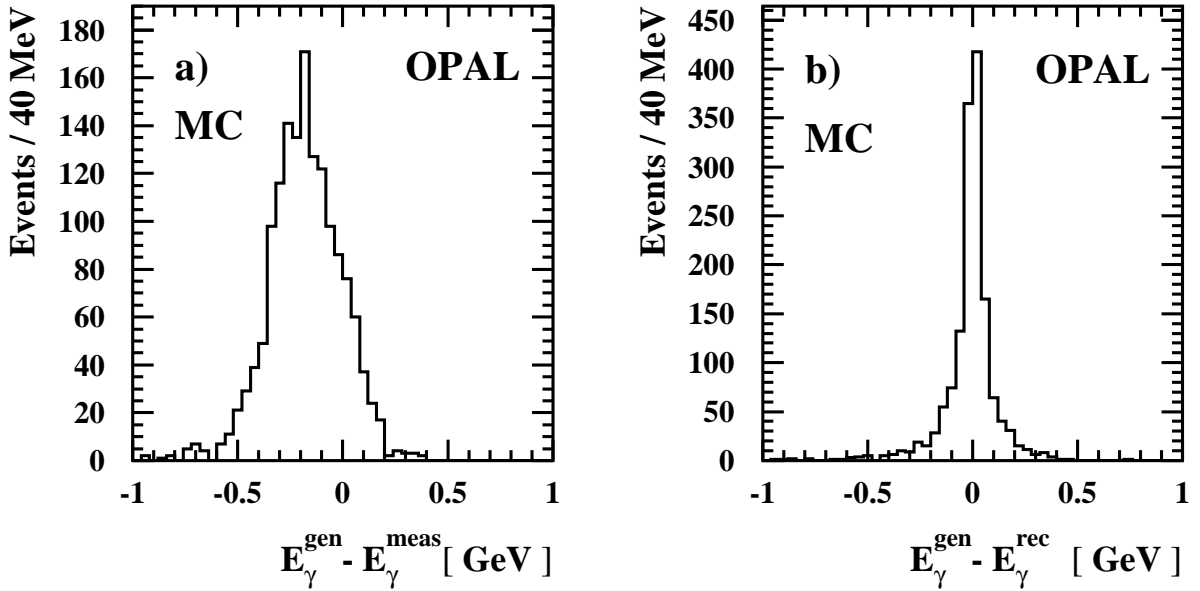} 
           }
   \end{center}
\caption{ (a) Difference between the generated  and the measured photon energy,
         $E_{\gamma}^{\rm gen}-E_{\gamma}^{\rm meas}$, for Monte Carlo (MC) 
         events after applying the cuts described in Section~\ref{sec-evsel}.
         (b) Difference between the generated and the reconstructed
         photon energy, $E_{\gamma}^{\rm gen}-E_{\gamma}^{\rm rec}$, 
         for the same Monte Carlo events.}
\label{fig-egam2}
\end{figure}

For signal events, 
the photon is one of the decay particles in a two-body decay, and so
its energy is kinematically fixed to be 430 MeV in the $\Chi$ rest frame.
Due to the Lorentz boost of the $\gg$ system, the energy of most photons 
is higher in the laboratory system but still often below 1 GeV. 
In Figure~\ref{fig-egam2}a, the
difference between the generated and the measured photon energy
after the detector response,
$E_{\gamma}^{\rm gen}-E_{\gamma}^{\rm meas}$, is shown for Monte Carlo  
events after applying the cuts described in Section~\ref{sec-evsel}.
In this regime the  energy resolution of the ECAL
does not allow a sufficiently precise measurement of the photon energy.

From momentum conservation,  
the transverse momentum of the photon
is balanced by the transverse momentum sum $\ptj$ of the two tracks. 
For the signal photon candidates, the energy and momentum 
is therefore reconstructed using the relation
\begin{equation}
E_{\gamma}^{\rm rec} = \frac{\ptj}{\sin\theta_{\gamma}},
\label{eq-egamma}
\end{equation}
with $\theta_{\gamma}$ being the polar angle of the ECAL cluster 
assumed to originate from the photon. 
This procedure improves the energy resolution significantly. 
The distribution of the difference between the generated and the reconstructed
photon energy, $E_{\gamma}^{\rm gen}-E_{\gamma}^{\rm rec}$, 
is significantly narrower (Figure~\ref{fig-egam2}b).

\begin{figure}[htbp]
   \begin{center}
      \mbox{
          \epsfxsize=9.0cm
          \epsffile{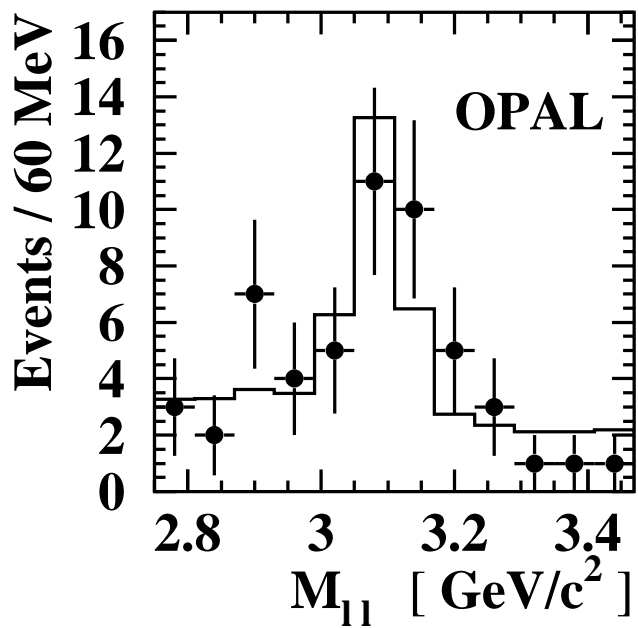}
           }
   \end{center}
\caption{Invariant mass $M_{\ell\ell}$ of the lepton-lepton system  after
         applying the set of cuts as explained in the text and after
         requiring the invariant mass of the $\ell\ell\gamma$ system 
         to be in the range $3.40<M_{\ell\ell\gamma}<3.65$ GeV$/c^2$ 
         around $M_{\Chi}$. A peak is visible at the mass of the $J/\psi$.
         The dots represent the data, while the histogram 
         shows the normalised Monte Carlo 
         on top of the background determined from data.  
         The Monte Carlo
         shows a small tail towards lower invariant masses $M_{\ell\ell}$ due
         to bremsstrahlung and final state radiation of the leptons.}
\label{jpsi-mass}
\end{figure}

Figure~\ref{jpsi-mass} shows the invariant mass spectrum of the two tracks 
after applying  all cuts and after
requiring the invariant mass of the lepton-lepton-photon  
($\ell\ell\gamma$) system 
to lie in the range $3.40<M_{\ell\ell\gamma}<3.65$~GeV$/c^2$ around $M_{\Chi}$.
The invariant mass $M_{\ell\ell\gamma}$ of
the $\ell\ell\gamma$ system is calculated
using the photon energy $E_{\gamma}^{\rm rec}$
reconstructed via Eq.~\ref{eq-egamma}.
A significant peak is visible around the mass of the $J/\psi$ 
at 3.1 GeV/$c^2$. The Monte Carlo simulation, represented by the histogram,
shows a small tail towards lower invariant masses. This is due
to bremsstrahlung and final state radiation of the leptons.

For a $\Chi$ event the invariant mass of the two tracks
must be the mass of the $J/\psi$.
Therefore a constrained fit is performed using a
program by Blobel~\cite{blobel}. 
For each track, the curvature, the azimuthal angle, 
the polar angle in the form 
$\cot\theta$ and the complete covariance matrix 
are used in the fitting algorithm.
From the fit we obtain a $\chi^2$ probability ${\cal P}(J/\psi)$
for the $J/\psi$ hypothesis.

\begin{figure}[htbp]
\centering
\unitlength1.cm
\begin{picture}(16,16)
\put(-1.,0.){ \epsfxsize=500pt \epsfbox{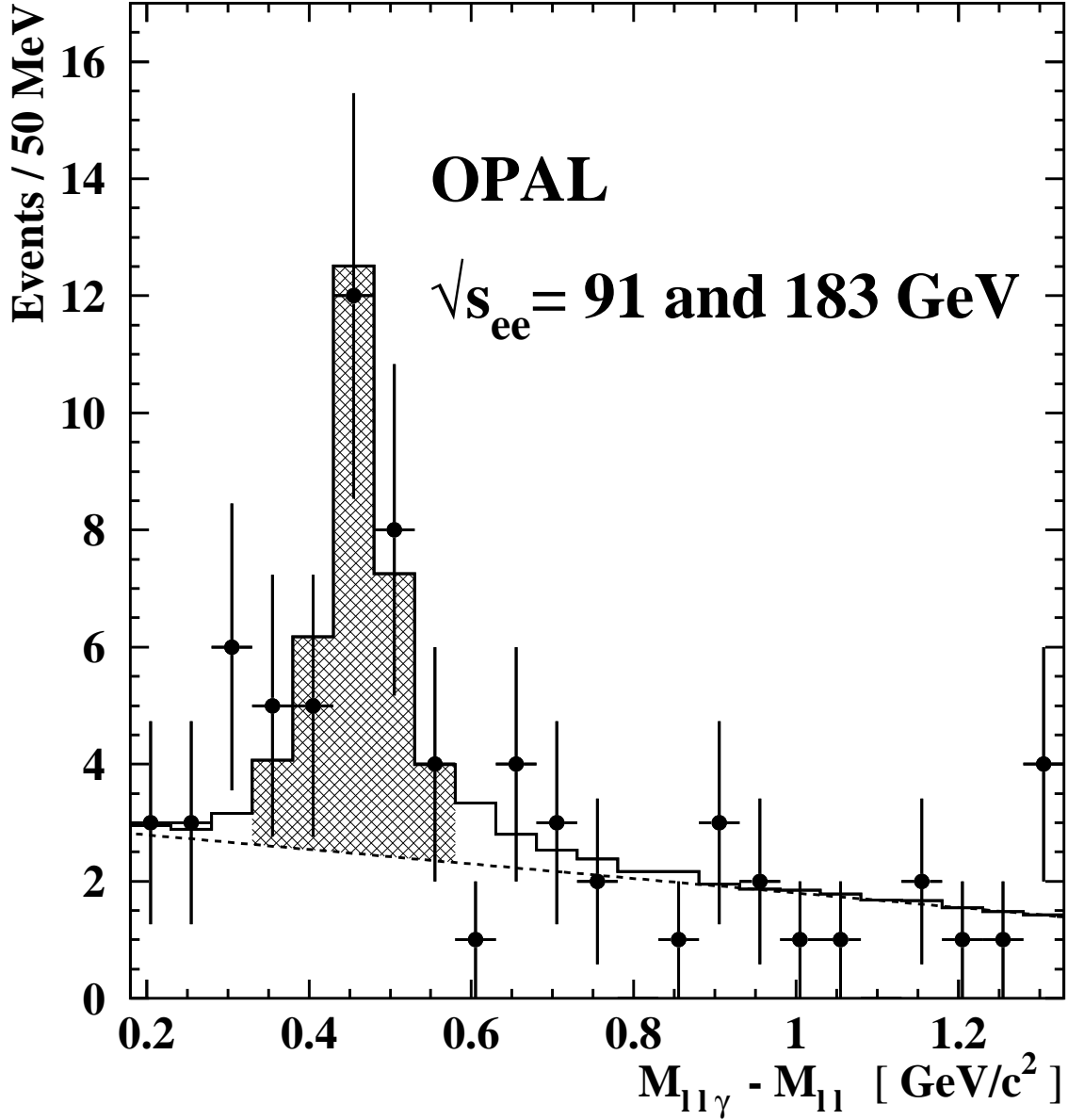} }
\end{picture}
\caption{ \rm Mass difference between the $\ell\ell\gamma$ and the
            $\ell\ell$ invariant mass, $M_{\ell\ell\gamma}-M_{\ell\ell}$, 
            after 
            applying the set of cuts explained in the text and after 
            requiring 
            ${\cal P} (J/\psi) > 1\%$.  A clear
            peak is visible around $M_{\Chi} - M_{J/\psi}= 459$~MeV$/c^2$.
            The fit of a linear function    
            to the sidebands of the signal is
            superimposed as dashed line. 
            The open histogram shows the normalised Monte
            Carlo added to the fitted background. The hatched area 
            represents the 
            invariant mass region between $330 < \DM < 580$~MeV$/c^2$.    }
\label{chi-mass}
\end{figure}

The mass difference between the $\ell\ell\gamma$ and the $\ell\ell$
system, $\DM$, is used to study the $\Chi$ signal. This minimizes
effects from the broadening of
the $\ell\ell \gamma$ and the $\ell\ell$ mass spectrum 
due to final state radiation of the leptons. 
The distribution of the mass difference, $\DM$, 
is shown in Fig.~\ref{chi-mass} for ${\cal P}(J/\psi) > 1 \%$.
A significant peak shows up around $\DM=459$~MeV$/c^2$ with 
$N_{\ell\ell\gamma}^{\rm SEL} = 34$ events counted in the signal region between
$330 < \DM < 580$~MeV$/c^2$.
The two tracks are identified as electrons
in 11 of the 34 selected events 
and in 23 events as muons. This
is consistent with the ratio of the Monte Carlo selection efficiencies 
for $J/\psi$ decays to electrons and muons, respectively.
Furthermore,     
7 of the 34 events are selected at $\sqee=183$~GeV which
is consistent with the ratio of $\Chi$ events expected for
the two $\ee$ centre-of-mass energies.
Fitting a Gaussian with a power tail plus a linear background function
to the data, where the exponent of the power tail
was determined from the Monte Carlo, yields a mean value
of $\langle\DM\rangle=464\pm14$~MeV$/c^2$ and a width of $39\pm16$~MeV$/c^2$
compared to $\langle\DM\rangle=459\pm1$~MeV$/c^2$ and a width of
$37\pm1$~MeV$/c^2$ in the Monte Carlo.

The two-photon width $\Gam$ is calculated from the number of 
events in the signal region, $N_{\ell\ell\gamma}^{\rm SEL}$, 
and the number of background
events in the signal region, $N_{\ell\ell\gamma}^{\rm BG}$, with
\begin{equation}
\Gam=\frac{1}{(2J_{\Chi}+1)}\;\frac{M_{\Chi}^3}{8\alpha_{\rm em}^2}\; 
\frac{1}{\BR(\Chi \to J/\psi~ \gamma)}\;
\frac{1}{\BR(J/\psi \to \ell^+\ell^-)}\;
\frac{N_{\ell\ell\gamma}^{\rm SEL}-N_{\ell\ell\gamma}^{\rm BG}}{
\sum g(z) L \epsilon_{\Chi}},
\label{eq-gam}
\end{equation}
where the sum runs over the two $\ee$ centre-of-mass energies $\sqee$.
The function $g(z)$ is defined by Eq.~\ref{eq-fz}. 
The total integrated luminosities for $\sqee=91$ and $183$~GeV are
$L=167$~pb$^{-1}$ and $55$~pb$^{-1}$. 
The selection efficiency is determined by dividing  
the sum of the weights of the
selected Monte Carlo events by the sum of the weights
of all generated events. 
The selection efficiencies $\epsilon_{\Chi}$ are $7.0 \%$ for  $\sqee=91$~GeV
and $5.7 \%$ for $\sqee=183$~GeV.  
Signal losses at the trigger level are found to be negligible.
The branching ratios
are $\BR(\Chi \to J/\psi~ \gamma)
= 0.135 \pm 0.011$ and
$\BR(J/\psi \to \ell^+\ell^-) = 0.1203 \pm 0.0027$~\cite{bib-pdg}.

The behaviour of the background is studied using
the complementary $\DM$ distribution with 
${\cal P}(\rm J/\psi) < 1 \%$. Its shape is
well described by a linear function. 
A linear function is therefore fitted to the sidebands of the $\DM$
spectrum with ${\cal P}(J/\psi)> 1 \%$ in Fig.~\ref{chi-mass}. 
The sidebands are defined by $\DM<280$~MeV$/c^2$ and $\DM>630$~MeV$/c^2$.
For the fit, a maximum likelihood method for Poisson-distributed data 
was used ~\cite{bib-pdg}.
The fit, which is superimposed, yields $N_{\ell\ell\gamma}^{\rm BG}=12.4\pm3.3$
background events in the signal region. 
The $\Chi$ Monte Carlo signal is added to the fitted background
after normalising it to the number of signal
minus background events in the signal region,
$N_{\ell\ell\gamma}^{\rm SEL}-N_{\ell\ell\gamma}^{\rm BG}$, of the data. 

There could be a small contribution to the $\Chi$ signal from 
$\ee\to\ee\chi_{\rm c0}$ and $\ee\to\ee\chi_{\rm c1} $ events
with $\chi_{\rm c0}$ and $\chi_{\rm c1} $ also decaying to $J/\psi~ \gamma$.
These possible contributions would increase the measured 
two-photon width $\Gam$ with respect to the actual width.
The mass peak of the $\chi_{\rm c0}$ meson is expected
to be at $\DM= 318.2\pm1.0$~MeV$/c^2$ compared to
$459.2\pm0.2$~MeV$/c^2$ expected for the $\Chi$~\cite{bib-pdg}.
Fig.~\ref{chi-mass} shows no indication for  
such a peak within the statistical errors.

\begin{figure}[htbp]
\centering
\vspace{3.cm}
\unitlength1.cm
\begin{picture}(7,4)
\put(-2.,0.){ \epsfxsize=250pt  \epsfbox{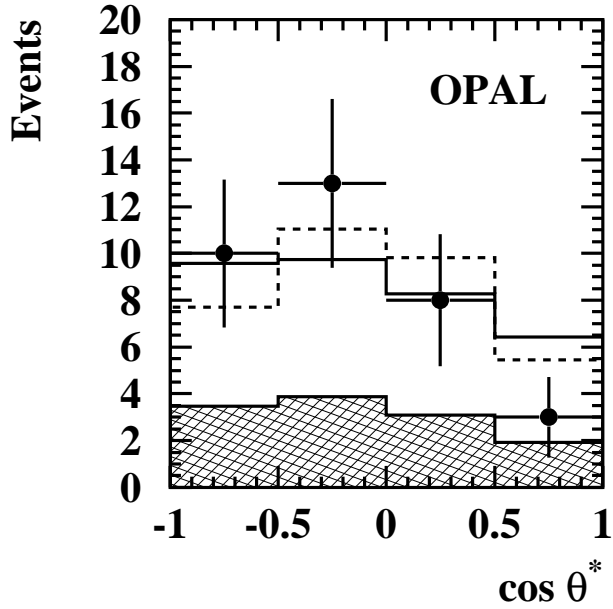} }
\end{picture}
\caption{
Cosine of the polar angle $\theta^{\ast}$ 
of the decay photon in the $\Chi$ rest frame.
The dots represent the data with their statistical
error, the hatched histogram represents the background taken from the sidebands
and then normalised to the number 
of background events in the signal region, $N_{\ell\ell\gamma}^{\rm BG}$.
The full line shows the sum of the normalised background and the 
signal Monte Carlo with the decay angular distribution  
as described in Section~\protect\ref{sec-MC}
after normalisation to the number of signal minus background events,
$N_{\ell\ell\gamma}^{\rm SEL}-N_{\ell\ell\gamma}^{\rm BG}$.
The dashed line is calculated in the same way, but for 
an isotropic decay angular distribution (spin 0).}
\label{angle}
\end{figure}

In principle, the contribution from $\chi_{\rm c0}$ decays
could be estimated from the data by measuring the decay angular
distribution. 
In Fig.~\ref{angle} the distribution of the polar angle of the photon, 
$\cos\theta^{\ast}$, is shown. The Monte Carlo expectation
for an isotropic decay ($J=0$) and for a decay angle
distribution as described in Section~\ref{sec-MC} for $J=2$ are superimposed.  
The agreement between data and Monte Carlo is reasonable in both cases.
Due to the small number of data events,
no further conclusion can be drawn concerning the decay angular 
distribution. 

The production cross-section of $\chi_{\rm c0}$ mesons
in two-photon events can
be calculated with Eq.~\ref{eq-ggres}. The ratio of the number of selected 
$\Chi$ events, $N_{\Chi}$,
to the number of selected $\chi_{\rm c0} $ events, $N_{\chi_{\rm c0}}$, 
is given by:
\begin{eqnarray}
\label{eq-ratio1}
\frac{N_{\Chi}}{N_{\chi_{\rm c0}}} 
                          & = &  \frac{\rm \sigma(\ee \to\ee \Chi)}
                            {\rm \sigma(\ee \to \ee \chi_{\rm c0})}\cdot
                                      \frac{\rm BR(\Chi \to J/\psi~ \gamma)}
                                           {\rm BR(\chi_{\rm c0} \to J/\psi~ \gamma)}\cdot
                                       \frac{\epsilon_{\Chi}}
                                            {\epsilon_{\chi_{\rm c0}}} \\
                   & = & \frac{2J_{\chi_{\rm c2}}+1}{2J_{\chi_{\rm c0}}+1}\cdot
 \rule{0mm}{10mm} \left( \frac{M_{\chi_{\rm c0}}} 
                                                   {M_{\Chi}} \right)^3\cdot 
                                       \frac{\Gam}
                                            {\Gamma(\chi_{\rm c0}\to\gamma\gamma)}\cdot
                                       \frac{\rm BR(\Chi \to J/\psi~ \gamma)}
                                            {\rm BR(\chi_{\rm c0} \to J/\psi~ \gamma)}\cdot
                                       \frac{\epsilon_{\Chi}}
                                            {\epsilon_{\chi_{\rm c0}}}. 
\label{eq-ratio2}
\end{eqnarray}
The spin factor equals 5, the mass term is of the order one and the
branching ratio $\rm BR(\Chi \to J/\psi~ \gamma)$ is about 20 times
larger than the branching ratio 
$\rm BR(\chi_{\rm c0} \to J/\psi~ \gamma)=0.0066\pm0.0018$~\cite{bib-pdg}.
The same set of cuts as for $\Chi$
was applied to Monte Carlo $\chi_{\rm c0}$ events yielding
a selection efficiency $\epsilon_{\chi_{\rm c0}}= 2.8 \%$
for $330<\DM<580$~MeV$/c^2$ at $\sqee=91$~GeV.
The two-photon width for 
the $\chi_{\rm c0}$ has been measured to be 
$\Gamma(\chi_{\rm c0}\to\gamma\gamma)=5.4\pm3.7$~keV~\cite{bib-lee}.
With this value and our measurement of $\Gam$, the background due to 
$\chi_{\rm c0}$ events is estimated to be $1.3\%$. 
This contribution is therefore neglected.

The case of the $\chi_{\rm c1}$ meson is more complicated, although 
the production of spin-1 particles in the collisions of real photons is
not allowed due to the Landau-Yang theorem~\cite{landau}. But since
the incoming photons have $Q^2>0$, 
the production of $\chi_{\rm c1}$ mesons is possible. 
The $\chi_{\rm c1}$ meson can not be well distinguished experimentally
from the $\Chi$, since
their masses are almost degenerate. For the $\chi_{\rm c1}$ mesons, the nominal mass 
difference $\DM$ is $413.6\pm0.2$~MeV$/c^2$~\cite{bib-pdg}.
The selection efficiency for $\chi_{\rm c1}$ events is determined
using GALUGA which takes into account the $Q^2$
dependence for a spin-1 resonance. 
The selection efficiency is found to be much smaller, 
$\epsilon_{\chi_{\rm c1}} \approx 0.1\%$ at $\sqee=91$~GeV,
than for $\Chi$ production,
because the average $Q^2$ of the $\chi_{\rm c1}$ events
is much higher and the transverse momentum of the
$\chi_{\rm c1}$ decay particles is therefore not balanced in the
detector. Most of the $\chi_{\rm c1}$ events are rejected due to the
requirements that there should be no energy deposit in the FD or SW 
and that the angle in the $r\phi$ plane between the photon and
the $J/\psi$ candidate must be greater than $150^{\circ}$.
Using the ratio of cross-sections given by the
model of Schuler~\cite{schuler} and Eq.~\ref{eq-ratio1} for
$\chi_{\rm c1}$ instead of $\chi_{\rm c0}$ mesons
with $\BR(\chi_{\rm c1} \to J/\psi~\gamma)=0.273\pm0.016$~\cite{bib-pdg}, 
the contribution from $\chi_{\rm c1}$
events is estimated to be much smaller than $1 \%$.
Background from $\psi'$ production in $\ee$ annihilation events   
has been  estimated to be negligible.

Equation~\ref{eq-gam} yields a measured two-photon width of
$$\Gam=\result\mbox{~keV}.$$
The first error is statistical, the second is systematic and
the last error is due to the uncertainty of the branching
ratios in the decay $\Chi \to J/\psi~\gamma \to \ell\ell\gamma.$

\section{Systematic errors}
\begin{figure}[htbp]
   \begin{center}
      \mbox{
          \epsfxsize=17.0cm
          \epsffile{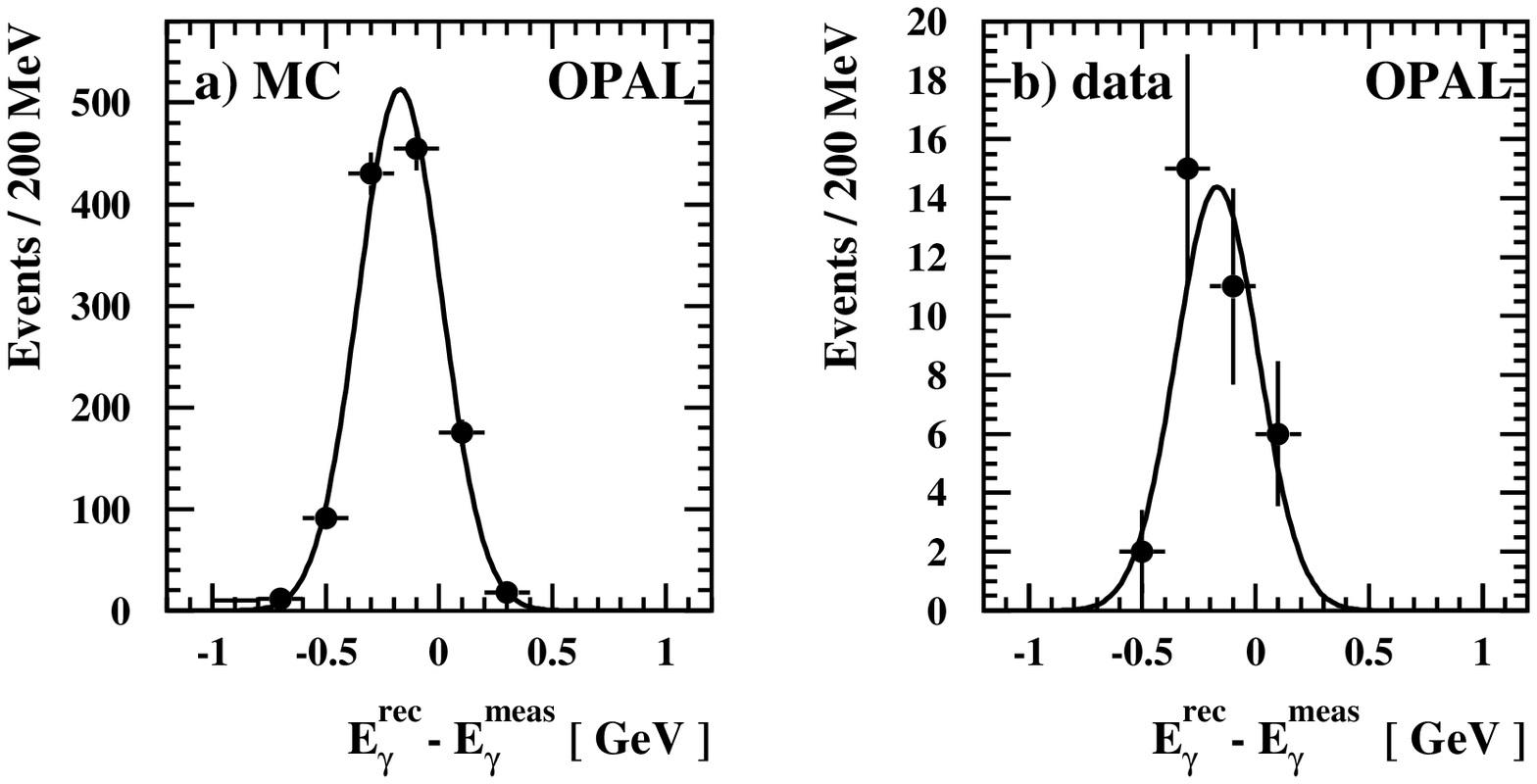} 
           }
   \end{center}
\caption{       Difference between the reconstructed and the measured
                photon energy, $E_{\gamma}^{\rm rec}-E_{\gamma}^{\rm meas}$, 
                for (a) the Monte Carlo (MC), and (b) the data, after
                all selection cuts. A Gaussian is fitted to both
                distributions yielding approximately the same mean and width,
                but with large errors in case of the data. }
\label{fig-egam}
\end{figure}

The following systematic errors are taken into account:
\begin{itemize}
\item The selection efficiency depends on the
$\Chi$ decay angular distribution used in the Monte Carlo generation.
In addition to the decay angular distribution as described in 
Section~\ref{sec-MC}
the Monte Carlo events are weighted according to a pure electromagnetic
dipole angular distribution (E1). This leads to a decrease of the
measured two-photon width $\Gam$ by $6 \%$.
\item 
The dependence on the form factor is studied using a $\rho$ form factor
and a $Q^2$ independent form factor
instead of the default $J/\psi$ form factor.
The selection efficiencies as well as the cross-sections calculated with
Eq.~\ref{eq-ggres} change, together contributing 
a $5 \%$ uncertainty to $\Gam$. 
\item
The error due to the cut on the ${\mathrm d}E/{\mathrm d}x$ probabilities
for the muon and 
electron hypotheses is estimated to be $4 \%$. 
\item
The minimum cluster energy of 300 MeV required in the ECAL for a 
photon candidate is an additional source of a systematic uncertainty. 
The energy resolution is given
by the width of the $E_{\gamma}^{\rm rec}-E_{\gamma}^{\rm meas}$ distribution
of the ECAL energy.
If the energy resolution of the Monte Carlo is better than the actual 
resolution in the experiment, 
the selection efficiency from 
the Monte Carlo would
be overestimated. To study this effect we exploit the fact that
the reconstructed photon energy $E_{\gamma}^{\rm rec}$ is approximately equal 
to the generated photon energy $E_{\gamma}^{\rm gen}$
for a $\Chi$ event. This is shown in Figure~\ref{fig-egam2}b.
The difference of the reconstructed and the measured photon
energy, $E_{\gamma}^{\rm rec}-E_{\gamma}^{\rm meas}$, is shown in 
Figure~\ref{fig-egam} for data and Monte Carlo after
all selection cuts. 
A Gaussian is fitted to both spectra yielding $-169 \pm 42$~MeV for
the mean in the data and $-173 \pm 5$~MeV for the mean in the Monte Carlo.
The widths are $181 \pm 34$~MeV for the data and $183 \pm 4$~MeV for 
the Monte Carlo. Mean and width are consistent within the statistical errors
of the fit.

The measured energy in the Monte Carlo is smeared with a Gaussian 
distribution in such a way that the decreased resolution
in the Monte Carlo is
approximately equal to the data resolution plus its error. 
This increases the measured $\Gam$ by $11 \%$.
\item
The error on the number of background events determined from
the fit of a linear function to the sidebands of the signal in the
mass difference spectrum  corresponds
to an additional uncertainty on $\Gam$  of $15 \%$. 
\item 
The statistical error of the 
Monte Carlo selection efficiency is $2 \%$.
\end{itemize}
The error of the luminosity measurement is negligible. 
The systematic errors due to the modelling of the energy resolution
and due to the background description are determined using
the data and therefore inevitably contain a considerable statistical 
component.
A summary of the errors is given in Table~\ref{errors}.
The errors are added quadratically yielding a total systematic
error of 21$\%$.

\section{Conclusions}
We have measured the two-photon width $\Gam$ in the process
$\ee\to\ee\Chi$, with the $\Chi$ mesons reconstructed in
the decay channel 
$\Chi \to  J/\psi ~\gamma \to  \ell^+~ \ell^- ~\gamma $ (with $\ell$ = $e, \mu$).
The whole OPAL data samples 
taken at $\sqee$
of $91$ and $183$~GeV are used, corresponding to  
$L=167$~pb$^{-1}$ and $55$~pb$^{-1}$.
In total we have selected 34 events in the $\Chi$ signal region
$330 < \DM < 580$~MeV$/c^2$ including a background of $12.4\pm3.3$ events. 
The contribution from $\chi_{\rm c0}$ and $\chi_{\rm c1}$ events 
is estimated not to exceed a few percent.
The two-photon width $\Gam$ is determined to be
$$\Gam=\result\mbox{~keV}.$$

The first error is statistical, the second is systematic and
the last error is due to the uncertainty of the branching
ratios in the decay $\Chi \to J/\psi~\gamma \to \ell\ell\gamma.$
In Table~\ref{tab-results}, a comparison 
of different measurements of the two-photon 
width $\Gam$ is given. Our result agrees with the results from CLEO,
TPC/2$\gamma$ and R704 if one takes into account the
large statistical and systematic errors. 
Our result is about two standard deviations
larger than the E760 result and the current world
average~\cite{bib-pdg} which is dominated by the E760 measurement.
It is also more than two standard deviations larger than the
prediction of Schuler~\cite{schuler}.

\vspace{-4mm}
\section*{Acknowledgements}
We thank Michael Sivertz for providing the calculation
of the decay angular distribution used by CLEO and Gerhard
Schuler for useful discussions. 
We particularly wish to thank the SL Division for the efficient operation
of the LEP accelerator at all energies
 and for their continuing close cooperation with
our experimental group.  We thank our colleagues from CEA, DAPNIA/SPP,
CE-Saclay for their efforts over the years on the time-of-flight and trigger
systems which we continue to use.  In addition to the support staff at our own
institutions we are pleased to acknowledge the  \\
Department of Energy, USA, \\
National Science Foundation, USA, \\
Particle Physics and Astronomy Research Council, UK, \\
Natural Sciences and Engineering Research Council, Canada, \\
Israel Science Foundation, administered by the Israel
Academy of Science and Humanities, \\
Minerva Gesellschaft, \\
Benoziyo Center for High Energy Physics,\\
Japanese Ministry of Education, Science and Culture (the
Monbusho) and a grant under the Monbusho International
Science Research Program,\\
German Israeli Bi-national Science Foundation (GIF), \\
Bundesministerium f\"ur Bildung, Wissenschaft,
Forschung und Technologie, Germany, \\
National Research Council of Canada, \\
Research Corporation, USA,\\
Hungarian Foundation for Scientific Research, OTKA T-016660,
T023793 and OTKA F-023259.\\

\begin{appendix}
\section*{Appendix}
\begin{figure}[htbp]
\vspace{-1.cm}
   \begin{center}
      \mbox{
          \epsfxsize=10.5cm
          \epsffile{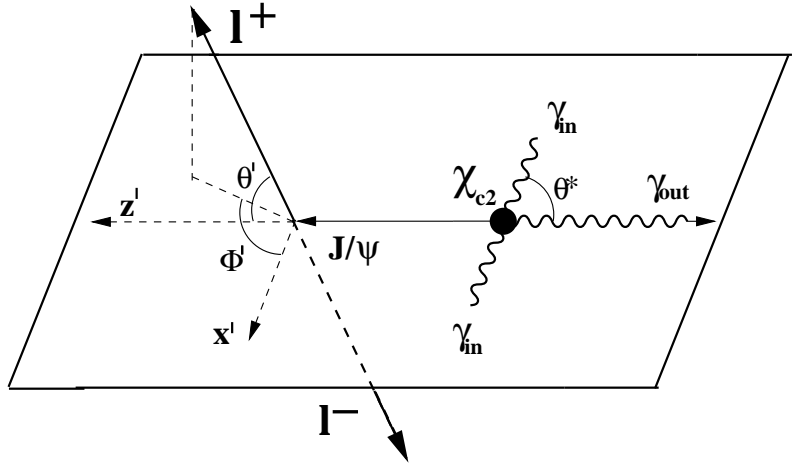}
           }
   \end{center}
\caption{Illustration of the $\Chi$ decay angles:
The polar angle $\theta^{\ast}$ is measured in the $\Chi$ rest frame 
and is the angle between the $\Chi$ direction of motion 
in the laboratory system
and the direction of the outgoing decay photon ($\gamma_{\rm out}$). 
The $\Chi$ direction of motion 
is collinear with the axis of the incoming photons ($\gamma_{\rm in}$). 
The polar and azimuthal angles of the positive decay lepton $\ell^+$,
denoted by $\theta^{'}$ and $\phi^{'}$, 
are measured in the $J/\psi$ rest frame with the $z'$ axis 
given by the direction opposite to
the photon's direction from the $\Chi$ decay,
and the $x'$ axis in the plane 
containing the $\Chi$ direction and the incident photons.}
\label{fig-angles}
\end{figure}

The angular distribution of the decay $\Chi \to J/\psi~\gamma \to
\ell^+\ell^-\gamma$ can be expressed in terms of the decay helicity
amplitudes $A_{|\nu'|}$~\cite{bib-angle}
\begin{eqnarray}
W(\theta^{\ast},\theta',\phi')
 \propto \sum_{\lambda=\pm2}
                                        \sum_{\nu=-2}^{2}
                                        \sum_{\nu^{'}=-2}^{2}
                                        \sum_{\mu=\pm1}
              A_{|\nu|}^{\ast}\;A_{|\nu^{'}|}\;
              d_{\nu \lambda}^{J=2}(\theta^{\ast})
d_{\nu^{'}\lambda }^{J=2}(\theta^{\ast})\;
              \rho^{\sigma\sigma^{'}}(\theta^{'},\phi^{'}),
\end{eqnarray}
where $\lambda$ describes the two possible helicities of the $\Chi$ with respect
to the $\gamma\gamma$ axis assuming that the $\Chi$ is
produced in a helicity $\pm2$ state.
The angles $\theta^{\ast},\theta'$ and
$\phi'$ are defined in Fig.~\ref{fig-angles}.
The indices $\nu$ and $\nu^{'}$ denote 
the helicity of the $\Chi$ with respect to the $J/\psi~\gamma$ axis and  
the index $\mu$ the two possible helicity states of the
outgoing photon. The definition of the functions $d_{\nu \lambda}^{J=2}$
can be found in Ref.~\cite{bib-pdg}. The angular
distribution of the $J/\psi$ decay 
is described by the density matrix $\rho^{\sigma\sigma^{'}}$,
where $\sigma=\nu-\mu$ and $\sigma^{'}=
\nu^{'}-\mu$ are the $J/\psi$ helicities with respect to
the $J/\psi~\gamma$ axis. Explicitly, 
the angular distribution of the decay $\Chi \to J/\psi~\gamma \to
\ell^+\ell^-\gamma$ is given by
\begin{eqnarray} 
\nonumber
\lefteqn{ W(\theta^{\ast},\theta',\phi') \propto
\frac{1}{8} A_2^2(1+\cos^2\theta^{'})
             (1+6\cos^2\theta^{\ast}+\cos^4\theta^{\ast})} \\
\nonumber
& & \pzz +A_1^2 (1-\cos^2\theta^{'})\;(1-\cos^4\theta^{\ast}) \\
\nonumber 
& & \pzz +\frac{3}{4} A_0^2(1+\cos^2\theta^{'})\;
             (1-2\cos^2\theta^{\ast}+\cos^4\theta^{\ast})  \\
\nonumber
& & \pzz + \frac{1}{2\sqrt2}A_2 A_1(\sin2\theta^{'}\cos2\phi^{'})
             (\sin\theta^{\ast}\cos\theta^{\ast}(3-\cos^2\theta^{\ast})) \\
\nonumber
& & \pzz +\frac{\sqrt6}{4}A_2 A_0 
           (\sin^2\theta^{'}\cos2\phi^{'})  
           (1-\cos^4\theta^{\ast}) \\
& & \pzz -\frac{\sqrt3}{2}\;A_1A_0  (\sin2\phi^{'}\cos2\phi^{'})
               (\sin\theta^{\ast}\cos\theta^{\ast}-
               \sin\theta^{\ast}\cos^3\theta^{\ast}).
\end{eqnarray}
The decay amplitudes have been determined
to be $A_2=0.85\pm0.03$, $A_1=0.49\pm0.03$ and $A_0=0.21\pm0.05$, 
based on measurements by 
E760~\cite{bib-E760a}. 
\end{appendix}

\clearpage
\begin{table}[htpb]
\begin{center}
\begin{tabular}{|l|c|c|}
\hline
\rule[-.25cm]{0.cm}{.8cm}Experiment         & $\;\,\Gam$ [keV]            
& measured in    \\ \hline\hline
\rule[-.25cm]{0.cm}{.8cm} CLEO~\cite{bib-cleo}      
&$1.08\pz\pm0.30\pz\pm0.26\pz$&$\ee\to\ee\Chi $   \\ \hline
\rule[-.25cm]{0.cm}{.8cm} TPC/2$\gamma$~\cite{bib-tpc}
&$3.4\pzz\pm1.7\pzz\pm0.9\pzz$   &$\ee\to\ee\Chi $  \\ \hline
\rule[-.25cm]{0.cm}{.8cm} E760~\cite{bib-E760b} 
&$0.321\pm0.078\pm0.054$&$\rm p\bar{p} \to \Chi \to \gamma\gamma$\\ \hline
\rule[-.25cm]{0.cm}{.8cm} R704~\cite{bib-r704}  
&$2.0\pzz\pm\,^{0.9}_{0.7}\pzz\pm 0.3\pzz$ & $\rm p\bar{p} \to\Chi\to \gamma\gamma$  \\ \hline \hline
\rule[-.25cm]{0.cm}{.8cm} average \cite{bib-pdg}  
&$0.37\pm0.17 $ 
&  \\ \hline
\rule[-.25cm]{0.cm}{.8cm} this measurement  
&$\result$& $\ee\to\ee\Chi $   \\ \hline

\end{tabular}
\caption{Results on the two-photon width $\Gam$. 
The first error is statistical, the second error systematic.
The CLEO and TPC$/2\gamma$ results are based on $25.4\pm6.9$ 
and $6.2\pm3.0$ signal events, respectively.
The R704 measurement is updated by
using the current world averages for the total width $\Gamma_{\rm tot}$
and for $\BR(\Chi\to\ppbar)$ from Ref.~\protect\cite{bib-pdg}.
It should be noted that R704 has used an isotropic angular
distribution for the decay $\Chi\to\gamma\gamma$ 
to calculate their efficiencies.}
\label{tab-results}
\end{center}
\end{table}
\vspace*{2.cm}

\begin{table}[htpb]
\begin{center}
\begin{tabular}{|c|c|}
\hline
\rule[-.25cm]{0.cm}{.8cm}source & relative error  \\ \hline\hline
\rule[-.25cm]{0.cm}{.8cm}$ \Chi$ decay angular distribution &  $ \pz6 \%$  \\ \hline
\rule[-.25cm]{0.cm}{.8cm}form factor                       &  $  \pz5 \%$   \\ \hline
\rule[-.25cm]{0.cm}{.8cm}energy loss of tracks             &  $  \pz4 \%$   \\ \hline
\rule[-.25cm]{0.cm}{.8cm}minimum cluster energy            &  $  11   \%$ \\ \hline
\rule[-.25cm]{0.cm}{.8cm}background determination          &  $  15   \%$   \\ \hline
\rule[-.25cm]{0.cm}{.8cm}Monte Carlo statistics            &  $  \pz2 \%$ \\ \hline\hline
quadratic sum $   \rule[-.5cm]{0.cm}{1.5cm}$ &$ 21 \% $ \\ \hline   
\end{tabular}
\vspace{.3cm}
\caption{ Summary of the systematic errors. The errors are added quadratically.}
\label{errors}
\end{center}
\end{table}
\vspace{-3.cm}

\newpage
\begin{thebibliography}{99}
\bibitem{bib-ppbar}
E760 Collaboration, T.A.~Armstrong et al., Phys.~Rev.~Lett.~68 (1992) 1468;\\
E760 Collaboration, T.A.~Armstrong et al., Nucl.~Phys.~B373 (1992) 35.
\bibitem{bodwin}
G.T.~Bodwin, E.~Braaten and G.P.~Lepage, Phys.~Rev.~D46 (1992) 1914.
\bibitem{schuler}
G.A.~Schuler, F.A.~Berends and R.~van~Gulik, 
{\em 
Meson-photon transition form factors and resonance
cross-sections in e$^+$e$^-$ collisions,} CERN-TH/97-294 (1997)
and hep-ph/9710462.
\bibitem{bib-cleo}
CLEO Collaboration, J.~Dominick et~al., Phys.~Rev.~D50 (1994) 4265.
\bibitem{bib-tpc}
TPC/$2\gamma$ Collaboration, D.A.~Bauer et~al., Phys.~Lett.~B302 (1993) 345. 
\bibitem{bib-E760b}
E760 Collaboration, T.A.~Armstrong et~al., Phys.~Rev.~Lett. 70 (1993) 2988.
\bibitem{bib-r704}
R704 Collaboration, C.~Baglin et al., Phys.~Lett.~B187 (1987) 191.
\bibitem{bib-pdg}
R.M.~Barnett et~al., Review of Particle Physics, Phys.~Rev.~D54 (1996) 1.
\bibitem{opaltechnicalpaper}
OPAL Collaboration, K.~Ahmet et~al., Nucl.~Instrum.~Methods~A305 (1991) 275;\\
P.P.~Allport et~al., Nucl.~Instrum.~Methods~A346 (1994) 476;\\
P.P.~Allport et~al., Nucl.~Instrum.~Methods~A324 (1993) 34;\\
O.~Biebel et~al., Nucl.~Instrum.~Methods~A323 (1992) 169;\\
M.~Hauschild et~al., Nucl.~Instrum.~Methods~A314 (1992) 74.
\bibitem{bib-siw}
B.E.~Anderson et~al., IEEE Transactions on Nuclear Science 41 (1994) 845.
\bibitem{bib-budnev}
V.M.~Budnev et~al., Phys.~Rep.~15C (1975) 181. 
\bibitem{bib-twogen} 
A.~Buijs, W.G.J.~Langeveld, M.H.~Lehto and D.J.~Miller, Comp.~Phys.~Comm.~79 (1994) 523.
\bibitem{bib-jetset}
T.~Sj\"ostrand, Comp.~Phys.~Comm.~82 (1994) 74;\\
T.~Sj\"ostrand, LUND University Report, LU-TP-95-20 (1995).
\bibitem{bib-gopal}
J.~Allison et~al., Nucl.~Instrum.~Methods A317 (1992) 47.
\bibitem{bib-close}
Z.P.~Li, F.E.~Close, T.~Barnes, Phys.~Rev.~D43 (1991) 2161.
\bibitem{bib-galuga}
G.A.~Schuler, Comp.~Phys.~Comm.~108 (1998) 279.
\bibitem{blobel} 
V.~Blobel, {\em Constrained Least Squares and Error Propagation}, 
to be published, Hamburg (1997).
\bibitem{bib-lee}
R.A.~Lee, PhD thesis, SLAC-0282, quoted in Ref.~\protect\cite{bib-pdg}. 
\bibitem{landau}
C.N.~Yang, Phys.~Rev. 77 (1950) 242.
\bibitem{bib-angle}
M.G.~Olsson and Casimir~J.~Suchyta III, Phys.~Rev.~D34 (1986) 2043.
\bibitem{bib-E760a}
E760 Collaboration, T.A.~Armstrong et~al., Phys.~Rev.~D48 (1993) 3037.
\end{thebibliography}
\end{document}